\documentclass[aps,pra,showpacs, twocolumn,amsmath,amssymb,tightenlines,epsfig,floatfix]{revtex4-1}
\usepackage{graphicx}
\usepackage{dcolumn}	
\usepackage{bm}			
\usepackage{amsfonts}
\usepackage{xspace}
\usepackage{color}
\usepackage{epstopdf}

\usepackage{multirow}
\usepackage{array}

\begin{document}

\newcommand{\psihat}{\ensuremath{\hat{\psi}}\xspace}
\newcommand{\psihatd}{\ensuremath{\hat{\psi}^{\dagger}}\xspace}
\newcommand{\ahat}{\ensuremath{\hat{a}}\xspace}
\newcommand{\Ham}{\ensuremath{\mathcal{H}}\xspace}
\newcommand{\ahatd}{\ensuremath{\hat{a}^{\dagger}}\xspace}
\newcommand{\bhat}{\ensuremath{\hat{b}}\xspace}
\newcommand{\bhatd}{\ensuremath{\hat{b}^{\dagger}}\xspace}
\newcommand{\boldr}{\ensuremath{\mathbf{r}}\xspace}
\newcommand{\dr}{\ensuremath{\,d\mathbf{r}}\xspace}
\newcommand{\dk}{\ensuremath{\,d^3\mathbf{k}}\xspace}
\newcommand{\etal}{\emph{et al.\/}\xspace}
\newcommand{\ie}{i.e.\:}
\newcommand{\eq}[1]{Eq.\,(\ref{#1})\xspace}
\newcommand{\fig}[1]{Figure\,(\ref{#1})\xspace}
\newcommand{\abs}[1]{\left| #1 \right|} 
\newcommand{\proj}[2]{\left| #1 \rangle\langle #2\right| \xspace} 
\newcommand{\Qhat}{\ensuremath{\hat{Q}}\xspace}
\newcommand{\Qhatd}{\ensuremath{\hat{Q}^\dag}\xspace}
\newcommand{\phihatd}{\ensuremath{\hat{\phi}^{\dagger}}\xspace}
\newcommand{\phihat}{\ensuremath{\hat{\phi}}\xspace}
\newcommand{\boldk}{\ensuremath{\mathbf{k}}\xspace}
\newcommand{\boldp}{\ensuremath{\mathbf{p}}\xspace}
\newcommand{\boldsigma}{\ensuremath{\boldsymbol\sigma}\xspace}
\newcommand{\boldalpha}{\ensuremath{\boldsymbol\alpha}\xspace}
\newcommand{\grad}{\ensuremath{\boldsymbol\nabla}\xspace}
\newcommand{\parti}[2]{\frac{ \partial #1}{\partial #2} \xspace}
 \newcommand{\vs}[1]{\ensuremath{\boldsymbol{#1}}\xspace}
\renewcommand{\v}[1]{\ensuremath{\mathbf{#1}}\xspace}
\newcommand{\Psihat}{\ensuremath{\hat{\Psi}}\xspace}
\newcommand{\Psihatd}{\ensuremath{\hat{\Psi}^{\dagger}}\xspace}
\newcommand{\Vhatd}{\ensuremath{\hat{V}^{\dagger}}\xspace}
\newcommand{\Xhat}{\ensuremath{\hat{X}}\xspace}
\newcommand{\Xhatd}{\ensuremath{\hat{X}^{\dag}}\xspace}
\newcommand{\Yhat}{\ensuremath{\hat{Y}}\xspace}
\newcommand{\Yhatd}{\ensuremath{\hat{Y}^{\dag}}
\xspace}
\newcommand{\ddt}{\ensuremath{\frac{d}{dt}}
\xspace}
\newcommand{\nset}{\ensuremath{n_1, n_2,\dots, n_k}
\xspace}
\newcommand{\sah}[1]{{\textcolor{red}{#1}}}

\title{Heisenberg-limited metrology with a squeezed vacuum state, three-mode mixing and information recycling}
\author{Behnam Tonekaboni$^{1,*}$, Simon A. Haine$^{1}$, and Stuart S. Szigeti$^{1,2}$}
\affiliation{$^1$School of Mathematics and Physics,  University of Queensland, Brisbane, QLD, 4072, Australia}
\affiliation{$^2$ARC Centre for Engineered Quantum Systems, The University of Queensland, Brisbane, QLD 4072, Australia}
\email{uqbtonek@uq.edu.au}


\begin{abstract}
We have previously shown that quantum-enhanced atom interferometry can be achieved by mapping the quantum state of squeezed optical vacuum to one of the atomic inputs via a beamsplitter-like process [Phys.~Rev.~A \textbf{90}, 063630 (2014)]. Here we ask the question: is a better phase sensitivity possible if the quantum state transfer (QST) is described by a three-mode-mixing model, rather than a beamsplitter? The answer is yes, but only if the portion of the optical state not transferred to the atoms is incorporated via information recycling. Surprisingly, our scheme gives a better sensitivity for lower QST efficiencies, and with a sufficiently large degree of squeezing can attain near-Heisenberg-limited sensitivities for arbitrarily small QST efficiencies. Furthermore, we use the quantum Fisher information to demonstrate the near-optimality of our scheme.

\end{abstract}


\maketitle

\section{Introduction}

Squeezed light is a resource that enables the detection of phase shifts below the standard quantum limit (SQL) $1/\sqrt{N}$, where $N$ is the number of detected photons \cite{Caves:1981}. Famously, it was recently used to enhance the sensitivity of the GEO 600 gravitational-wave detector \cite{LIGO:2011} and the Hanford LIGO detector \cite{LIGO:2013}; this speaks to the maturity of squeezed-light-generation technology. As a consequence, squeezed light is a promising controllable resource for generating squeezed \emph{atomic} states  - and therefore enabling sub-SQL \emph{atom interferometry} - via a quantum state transfer (QST) process that maps the state of the squeezed light to the atomic field \cite{Kuzmich:1997, Hald:1999, Jing:2000, Fleischhauer:2002b, Haine:2005, Haine:2005b, Haine:2006b, Hammerer:2010}. Given the low fluxes of current atomic sources (compared with photon sources), and the technical barriers to increasing this flux \cite{Szigeti:2012, Robins:2013}, it is likely that atom interferometers could only measure potential violations of the weak equivalence principle \cite{Fray:2004, Dimopoulos:2007, Schlippert:2014}, quantum gravitational effects \cite{Amelino-Camelia:2009}, and gravitational waves \cite{Tino:2007, Dimopoulos:2008} if practical operation below the SQL is achieved.

Unfortunately, in practice QST between atoms and light is imperfect; any transmitted component of the optical squeezed state behaves as a loss mechanism that can drastically reduce the effectiveness of the squeezing \cite{Gea-Banacloche:1987, Demkowicz-Dobrzanski:2012}. However, the deleterious effects of imperfect QST can be reduced if the transmitted photons are instead measured, and the information obtained is combined in just the right way with the atom-interferometer signal. This technique is called \emph{information recycling} \cite{Haine:2013}, and we have previously shown its effectiveness in atom interferometry schemes enhanced by either single-mode or two-mode squeezed optical vacuum \cite{Szigeti:2014b}, and also for a general class of Heisenberg-limited input states \cite{Haine:2014b}. 

The majority of this previous work assumed that the QST process behaved as an \emph{atom-light beamsplitter} with a beamsplitter reflection coefficient $\mathcal{Q}$ corresponding to the efficiency of the QST process (e.g., when $\mathcal{Q} = 1$ all photons are mapped to atoms, when $\mathcal{Q} = 1/2$ only $50\%$ of the photons are mapped to atoms, etc.). This description is only valid when the mean number of photons in the input optical state is much smaller than the total number of atoms. In contrast, when the number of photons and atoms are comparable, the QST process is more appropriately described as a \emph{three-mode mixing process} (with one photon mode and two atomic modes corresponding to the mode initially occupied and the mode into which atoms are outcoupled during QST). We have previously investigated this QST process in the regime of moderate squeezing, where there is a quantitative but not qualitative difference to a beamsplitter QST process \cite{Szigeti:2014b}.

In this paper, we consider this three-mode mixing regime in considerably more detail, including the large squeezing regimes where the number of squeezed photons is comparable to the number of atoms, yielding dynamics and sensitivities that are both quantitatively and qualitatively different to the beamsplitter model of QST. In particular, we show that if a single-mode squeezed optical vacuum is partially mapped to two atomic inputs of a Mach-Zehnder (MZ) interferometer, and the transmitted photons are monitored and information recycling is applied, then lower QST efficiencies can in fact lead to \emph{better} phase sensitivities. Indeed, for a sufficiently large squeezing parameter, we show that near-Heisenberg-limited phase sensitivities are obtainable for arbitrarily low QST efficiencies. Crucially, this result is somewhat different to \cite{Pezze:2008}, which showed that a MZ interferometer can achieve Heisenberg-limited sensitivities if the two inputs are a coherent state and a single-mode squeezed vacuum state. In our scheme, the three-mode mixing QST process results in some non-trivial interferometer input state that is entangled with the transmitted photons. We also demonstrate the near-optimality of our phase-estimation procedure by computing the quantum Fisher information.


\section{Theoretical model} \label{sec_theoretical_model}

\subsection{Atom-light QST process} \label{sec_QST}
\begin{figure}[!t]
\centering
\includegraphics[width=0.6\columnwidth]{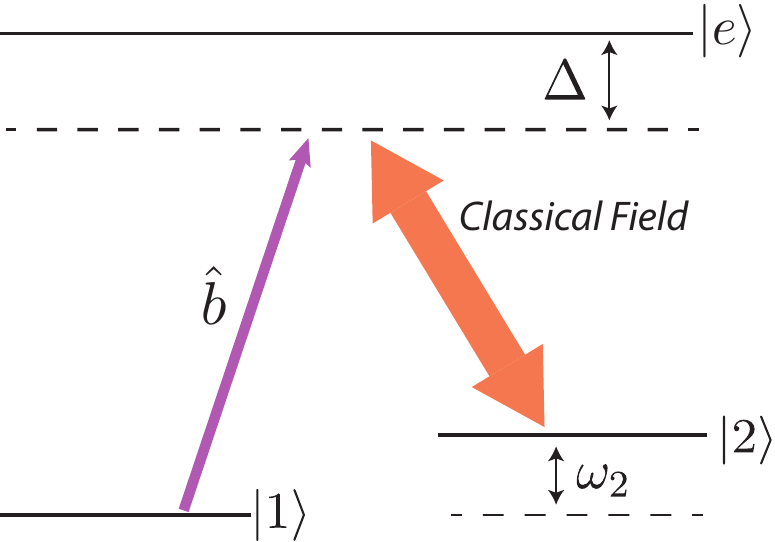}
\caption{ (Color online) Energy level scheme for a three-level Raman transition comprising two non-degenerate hyperfine ground states, $|1\rangle$ and $|2\rangle$. The BEC is initially formed in the state $|1\rangle$, and population is transferred to $|2\rangle$ via the absorption of a photon from mode $\hat{b}$ and the emission of a photon into the optical field addressing the $|e \rangle \leftrightarrow | 2 \rangle$ transition (which we treat as a large classical field). Since the detuning $\Delta$ of the optical fields is large, the excited state $| e \rangle$ is not appreciably populated on the timescale of the QST dynamics.}
\label{levelsfig}
\end{figure}

The simplified atom-light coupling model used throughout this paper is derived in considerable detail in \cite{Szigeti:2014b} (see also \cite{Haine:2005, Haine:2005b, Haine:2006b}), so we only briefly summarize it here. We consider an interactionless Bose-Einstein condensate (BEC) \footnote{For concreteness, we assume the atomic ensemble is Bose-condensed. However, many results in this paper equally apply to ultracold thermal vapors.} comprised of atoms with two hyperfine ground states $| 1 \rangle$ and $|2 \rangle$. These two states are coupled with a Raman transition via two counter-propagating optical fields, detuned from some excited state $| e \rangle$ that is not populated on the timescale of this coupling (see Fig.~\ref{levelsfig}). The optical field addressing the $| 2 \rangle \leftrightarrow |e \rangle$ transition is assumed to be a bright coherent state, and is therefore treated classically. Energy and momentum conservation ensures that the Raman transition has a high degree of mode selectivity. So, given a BEC that initially occupies a single motional mode of state $| 1 \rangle$ (mode $\hat{a}_1$), during a Raman transition predominantly one mode of the electromagnetic field addressing the $|1 \rangle \to |2 \rangle$ transition (mode $\hat{b}$) outcouples atoms to a different motional state in $|2 \rangle$ (mode $\hat{a}_2$). Here $\hat{a}_i$ and $\hat{b}$ are bosonic modes obeying the usual commutation relations. The effective interaction Hamiltonian describing this process is
\begin{equation}
	\hat{\mathcal{H}} =  \hbar g \left( \ahat_1 \ahatd_2 \bhat + H.c. \right), \label{simple_ham_3mode}
\end{equation}
where $g$ is the coupling strength between light and atoms, and we have assumed that the system is on-resonance such that the energy transferred by the two-photon transition perfectly matches the change in electronic and kinetic energies of the atom. 

Typically, the occupation of modes $\hat{a}_2$ and $\hat{b}$ is assumed to be much less than the occupation of mode $\hat{a}_1$. In this case we can treat $\hat{a}_1$ as an undepletable reservoir and make the approximation $\hat{a}_1 \to \sqrt{N_{a_1}}$, where $N_{a_1}$ is the mean number of atoms in mode $\hat{a}_1$. Consequently, Eq.~(\ref{simple_ham_3mode}) reduces to the beamsplitter Hamiltonian
\begin{equation}
	\hat{\mathcal{H}}_\text{BS} =  \hbar g \sqrt{N_{a_1}} \left( \ahatd_2 \bhat  + H.c. \right), \label{ham_BS}
\end{equation}
which corresponds to the linear evolution
\begin{subequations}
\label{analytic_QST}
\begin{align}
	\ahat_2(t) 		&= \sqrt{1 - \mathcal{Q}(t)} \, \ahat_2(t_0)  - i \sqrt{\mathcal{Q}(t)}\, \bhat(t_0), \label{asol} \\ 
	\bhat(t) 		&= \sqrt{1 - \mathcal{Q}(t)}\, \bhat(t_0)  - i \sqrt{\mathcal{Q}(t)}\, \ahat_2(t_0) \, \label{bsol}.
\end{align}
\end{subequations}
Here $\mathcal{Q}(t) = \sin^2(g \sqrt{N_{a_1}} t)$ is interpreted as the reflection coefficient of this atom-light beamsplitter, and therefore quantifies the efficiency of QST between the photon mode $\hat{b}$ and atomic mode $\hat{a}_1$. For example, if $\mathcal{Q}(t) = 1$ then the quantum state of $\hat{b}(t_0)$ is completely mapped to $\hat{a}(t)$, whereas if $\mathcal{Q}(t) = 0.5$ only 50\% of the photons couple to the atoms via the Raman transition.

When the number of photons is comparable with the number of atoms in the condensate, the undepleted reservoir approximation cannot be applied, and the QST process must be described by the three-mode mixing dynamics 
\begin{subequations}
\label{3modeEOM}
\begin{eqnarray}
i \dot{\ahat}_1 &=& g \ahat_2 \bhatd,  \label{ahat1dot} \\
i \dot{\ahat}_2 &=& g \ahat_1\bhat,  \label{ahat2dot} \\
i \dot{\bhat} &=& g \ahat_2\ahatd_1,  \label{bhatdot}
\end{eqnarray}
\end{subequations}
which are the Heisenberg equations of motion assuming Hamiltonian~(\ref{simple_ham_3mode}). Although an analytic solution is possible for times much shorter than $g^{-1}$ (see Sec.~\ref{sec_small_time_anal}), in general Eqs.~(\ref{3modeEOM}) must be solved numerically. 

Although beamsplitter dynamics no longer apply, the QST process can still be quantified via the \emph{generalized QST efficiency}
\begin{equation}
	\mathcal{Q}(t) \equiv \frac{\langle \hat{a}_2^\dag(t)\hat{a}_2(t) \rangle}{\langle \hat{b}^\dag(t_0)\hat{b}(t_0) \rangle}, \label{defn_Q} \\
\end{equation}
which compares the number of atoms outcoupled compared with the total number of input photons. 

This three-mode mixing process (\ref{3modeEOM}) has been previously studied within the context of non-degenerate parametric down-conversion \cite{Kinsler:1991, Drummond:1995}, and recently by us within the context of squeezed-light-enhanced atom interferometry \cite{Szigeti:2014b}. This prior analysis of three-mode mixing QST dynamics was in regimes which are qualitatively similar to a beamsplitter QST process. Below we considerably extend the analysis of the single-mode squeezed-light-enhanced atom interferometer presented in \cite{Szigeti:2014b}; in particular, we investigate the regime of moderate to large optical squeezing, where the three-mode mixing QST dynamics is qualitatively different to a beamsplitting QST process.

\begin{figure*}[!htb]
\centering
\includegraphics[width=\textwidth]{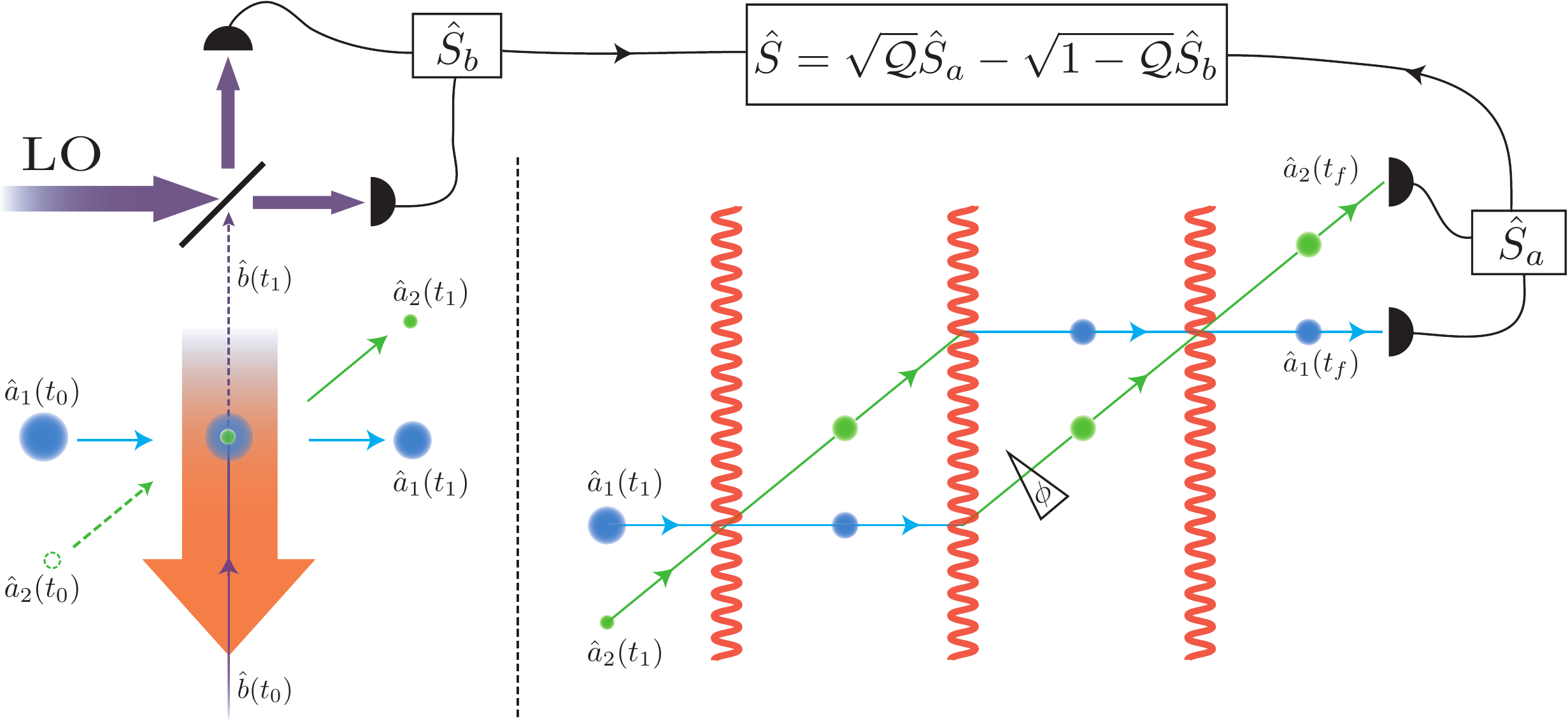}
\caption{ (Color online) A hybrid atomic-photonic interferometer enhanced with single-mode squeezed optical vacuum and information recycling. Left panel: Atoms in mode $\hat{a}_1$ are (partially) outcoupled to mode $\hat{a}_2$ via a stimulated Raman transition, with the goal of mapping the quantum state of the squeezed optical vacuum $\hat{b}(t_0)$ to $\hat{a}_2(t_1)$ (see Sec.~\ref{sec_QST}). Right panel: the two atomic modes $\hat{a}_1$ and $\hat{a}_2$ form the two arms of a MZ atom interferometer; any relative phase shift $\phi$ that accrues between the two modes during the evolution of the atoms through the interferometer can be determined by measuring the number difference, $\hat{\mathcal{S}}_a$, of the two atomic modes at time $t_f$. By interfering any light transmitted after the QST process [$\hat{b}(t_1)$] with a local oscillator (L.O.), assumed to be a large coherent state, a measurement of the phase quadrature $\hat{Y}_b$ of $\hat{b}(t_1)$ is achieved. Information recycling is then incorporated as a post-processing step, whereby the information-recycled signal $\hat{\mathcal{S}}$ is constructed from the atomic and photonic signals, $\hat{\mathcal{S}}_a$ and $\hat{\mathcal{S}}_b$ respectively. The signal $\hat{\mathcal{S}}$ generally gives a more sensitive phase measurement than the purely atomic signal $\hat{\mathcal{S}}_a$. }
\label{interferometer}
\end{figure*}

\subsection{Hybrid atomic-photonic interferometer with information recycling}

The atom-light QST process and information recycling is incorporated into a MZ interferometer as shown in Fig.~\ref{interferometer}. Initially (at time $t_0$), all the condensate atoms are in a coherent state in mode $\hat{a}_1$, and our photon mode is assumed to be a single-mode squeezed vacuum state \cite{Loudon:1987}
\begin{equation}
	\bhat(t_0) = \hat{\vartheta} \cosh r + \hat{\vartheta}^\dagger \sinh r,
\end{equation}
where $\hat{\vartheta}$ is a vacuum mode and $r > 0$ is the squeezing parameter. Our state is squeezed in the phase quadrature, as this is optimal for our scheme. Using an appropriately chosen coherent optical field, the optical mode $\hat{b}$ is coupled to modes $\hat{a}_1$ and $\hat{a}_2$ via a Raman transition, as described in Sec.~\ref{sec_QST}, with the aim of mapping the state of $\hat{b}(t_0)$ to $\hat{a}_2(t_1)$ (see left panel of Fig.~\ref{interferometer}). 

After the QST process is complete, the two atomic modes $\hat{a}_1(t_1)$ and $\hat{a}_2(t_1)$ form the inputs of a MZ atom interferometer (see right panel of Fig.~\ref{interferometer}). Here light pulses are used to coherently interfere, redirect, and recombine the two atomic modes. The evolution of the atomic modes through the interferometer is most conveniently described in terms of the pseudo-spin angular momentum operators:
\begin{subequations}
\label{J}
\begin{eqnarray}
	\hat{J}_x &= \frac12 (\ahat_2^\dagger \ahat_1 + \ahat_1^\dagger \ahat_2), \label{Jx}\\ 
	\hat{J}_y &= \frac i2 (\ahat_2^\dagger \ahat_1 - \ahat_1^\dagger \ahat_2), \label{Jy}\\
	\hat{J}_z &= \frac12 (\ahat_1^\dagger \ahat_1 - \ahat_2^\dagger \ahat_2). \label{Jz}\
\end{eqnarray}
\end{subequations} 
In particular, if the two modes accrued a relative phase shift $\phi$ during periods of free evolution within the interferometer, then this phase shift can be determined by measuring the relative number difference at time $t_f$:
\begin{equation}
	\hat{\mathcal{S}}_a \equiv 2 \hat{J}_z(t_f) = 2\left[  \hat{J}_z(t_1) \cos \phi - \hat{J}_x(t_1) \sin \phi \right].
\end{equation}
If there is complete QST, $\mathcal{Q} = 1$, then the minimum sensitivity of this phase measurement can surpass the SQL \cite{Caves:1981, Szigeti:2014b}:
\begin{equation}
	\Delta \phi = \frac{\sqrt{V(\hat{\mathcal{S}}_a)}}{ |\partial_\phi \langle \hat{\mathcal{S}_a} \rangle|} \approx \frac{e^{-r}}{\sqrt{N_t}} , \label{deltaphidef}
\end{equation}
where $N_t$ is the total number of atoms in the initial BEC, $V(\hat{O}) = \langle \hat{O}^2 \rangle - \langle \hat{O} \rangle^2$ is the variance, and we have assumed that $N_b \equiv \langle \hat{b}^\dag(t_0) \hat{b}(t_0) \rangle = \sinh^2 r \ll N_t$. 

However, for the typical case of $\mathcal{Q} < 1$, the squeezed optical state is only partially mapped to $\hat{a}_2$. Since, to first order, the QST process is a beamsplitter that mixes vacuum [$\hat{a}_2(t_0)$] with the squeezed optical input $\hat{b}(t_0)$ (i.e. a linear loss mechanism), the effects of the squeezing are drastically reduced, and the phase sensitivity is degraded. Fortunately, the lost portion of the metrologically useful quantum correlations still exist in the mode of the transmitted light, $\hat{b}(t_1)$. As discussed in \cite{Szigeti:2014b}, if the phase quadrature of the transmitted light $\hat{Y}_b = i[\hat{b}(t_1) - \hat{b}^\dag(t_1)]$ is measured via homodyne detection, then the \emph{information-recycled signal}
\begin{equation}
	\hat{\mathcal{S}} = \sqrt{\mathcal{Q}} \hat{\mathcal{S}}_a - \sqrt{1 - \mathcal{Q}} \hat{\mathcal{S}}_b, \label{eqn_info_recyc_sig}
\end{equation}
where $\hat{\mathcal{S}}_b = \sqrt{\smash[b]{\langle \hat{N}_{a_1}(t_1) \rangle}} \hat{Y}_b$, significantly ameliorates the effect of incomplete QST on the phase sensitivity. Again, the benefits of information recycling on this hybrid atom-photonic interferometer were only quantified for a) a beamsplitter QST process in the regime $\sinh^2 r \ll N_t$ and b) a three-mode mixing QST process in the regime of moderately-sized $r$ where the QST process behaves qualitatively as a beamsplitter. Below, we consider in detail the regime where $N_b \gtrsim N_t $, and show that information recycling gives surprisingly good sub-SQL phase sensitivities with a counterintuitive dependence on $r$ and $\mathcal{Q}$.


\section{Beamsplitter QST process} \label{Sec_undepleted}
Before considering the full three-mode mixing QST dynamics, we first consider the simplest case where the QST process is a beamsplitter [see Eqs.~(\ref{analytic_QST})]. Here, the phase sensitivity of the information-recycled signal $\hat{\mathcal{S}}$ can be computed analytically; although the beamsplitter QST model is invalid except in the regime where $\sinh^2 r \ll N_t$, this provides some basic intuition and develops a baseline to compare with the more counterintuitive results of Sec.~\ref{sec_single_mode_atom_int}. Specifically, at the optimal operating point $\phi = \pi/2$, the variance and slope of the signal $\hat{\mathcal{S}}$ are
\begin{subequations}
\begin{align}
	V(\hat{\mathcal{S}})	 &=  N_t e^{-2r}  + \mathcal{Q}(\mathcal{Q} - e^{-2r}) \sinh^2 r,  \label{varS}\\
	\partial_\phi \langle \hat{\mathcal{S}} \rangle &= \sqrt{\mathcal{Q}}(N_t - 2\mathcal{Q}\sinh^2r),
\end{align}
\end{subequations}
respectively, yielding a phase sensitivity of \cite{Szigeti:2014b}
\begin{equation}
\label{deltaphi_undepleted}
	\Delta \phi_\text{BS}(r, \mathcal{Q}) = \frac{\sqrt{\frac{N_t e^{-2r}}{\mathcal{Q}} + (\mathcal{Q} - e^{-2r})\sinh^2r}}{N_t-2\mathcal{Q}\sinh^2r}.
\end{equation}
In the limit $\sinh^2 r \ll N_t$, $\Delta \phi_\text{BS} \approx \exp(-r) / \sqrt{ \mathcal{Q} N_t}$, which surpasses the SQL provided $\mathcal{Q} > \exp(-2 r)$. However, in the intermediate regime where $r$ is large enough that $\mathcal{Q} \gg \exp(-2 r)$, and the second term under the square root in Eq.~(\ref{deltaphi_undepleted}) must be kept, but $N_t \gg 2 \mathcal{Q} \sinh^2 r$:
\begin{equation}
	\Delta \phi_\text{BS}(r, \mathcal{Q}) \approx \frac{\sqrt{\frac{N_t e^{-2r}}{\mathcal{Q}} + \mathcal{Q} \sinh^2r}}{N_t}. \label{approx_deltaphi_undepleted}
\end{equation}
Equation~(\ref{approx_deltaphi_undepleted}) attains a minimum of 
\begin{equation}
	\Delta \phi_\text{BS}^\text{min}(\mathcal{Q}) \approx \frac{\sqrt{N_t - \frac{\mathcal{Q}}{2}\sqrt{N_t} + \frac{\mathcal{Q}^2}{8}}}{N_t^{5/4}} \approx \frac{1}{N_t^{3/4}}, 
\end{equation}
at an optimal squeezing parameter of $r_\text{BS}^\text{opt}(\mathcal{Q}) \approx \ln(4N_t / \mathcal{Q}^2)/4$. 

\begin{figure}[tb]
\centering
\includegraphics[width=\columnwidth]{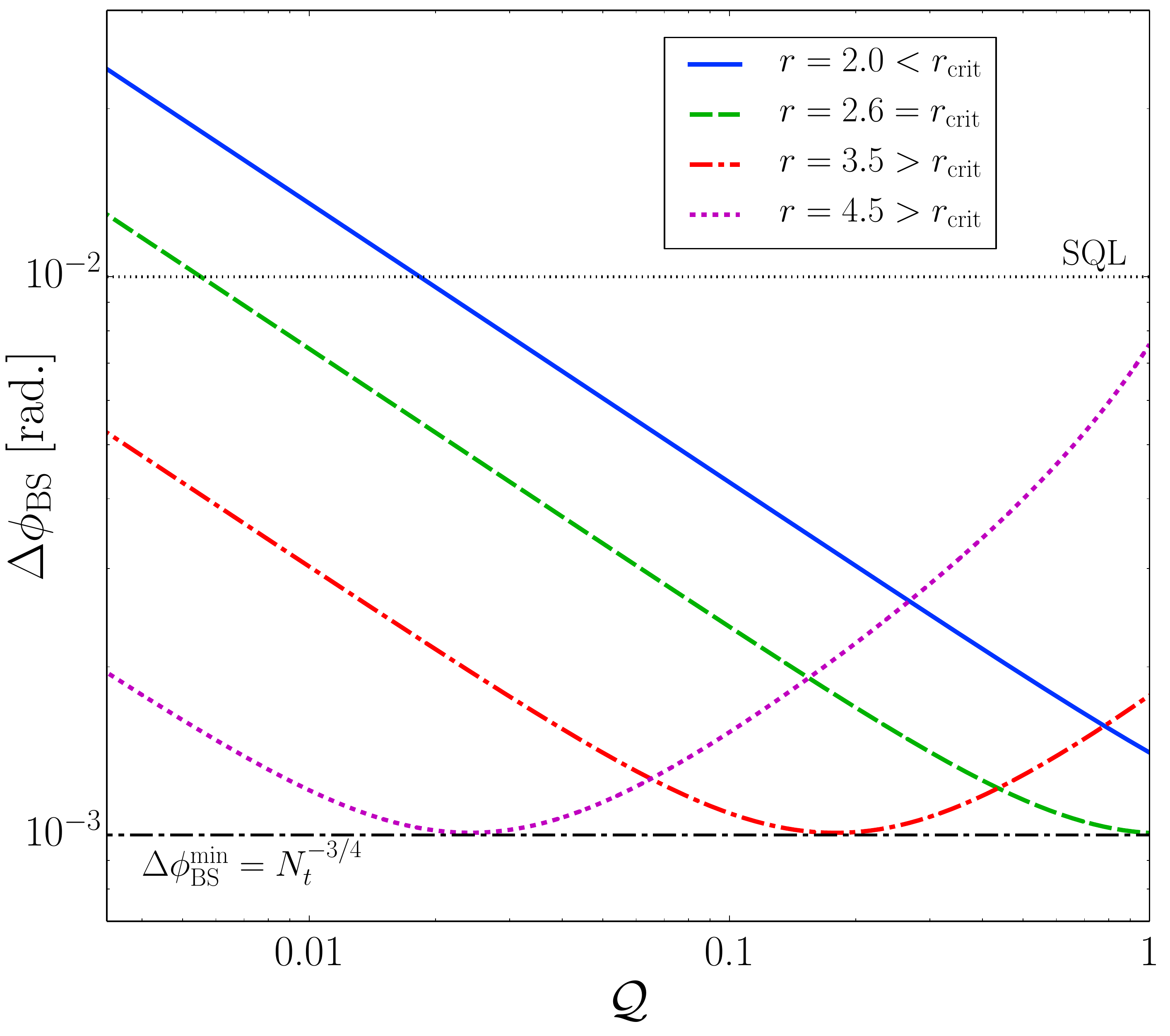}
\caption{(Color online) The QST-dependence of phase sensitivity Eq.~(\ref{deltaphi_undepleted}), which assumes a beamsplitter QST process, for different squeezing factors $r$ and an initial condensate of $N_t = 10^4$ atoms. The minimum phase sensitivity of $\Delta \phi_\text{BS}^\text{min} = N_t^{-3/4}$ is only attained if $r \geq r_\text{crit}$ and $\mathcal{Q} \approx 2 \exp(-2r) \sqrt{N_t}$.}
\label{fig_BS}
\end{figure}

We can therefore identify two qualitatively different regimes, delineated by $r_\text{crit} \equiv r_\text{BS}^\text{opt}(\mathcal{Q} = 1)$. When $r < r_\text{crit}$, the minimum sensitivity only occurs at $\mathcal{Q} = 1$, and is bounded by $N_t^{-3/4} < \Delta \phi_\text{BS}^\text{min} \leq \exp(-r) / \sqrt{N_t}$; generally this upper bound is a good approximation to $\Delta \phi_\text{BS}^\text{min}$. When $r > r_\text{crit}$ a minimum phase sensitivity of $\Delta \phi_\text{BS}^\text{min} \approx N_t^{-3/4}$ is \emph{always} possible, however this occurs at a QST fraction less than unity (see Fig.~\ref{fig_BS}). That the minimum phase sensitivity requires a \emph{reduction} in QST efficiency, and that furthermore $\mathcal{Q}_\text{opt} \approx 2 \exp(-2r) \sqrt{N_t} \to 0$ as $r$ gets large, is a curious feature of our hybrid atomic-photonic interferometer. However, for a beamsplitter QST process an imperfect QST efficiency does not give an improved sensitivity over the sensitivity at $\mathcal{Q} = 1$ and $r = r_\text{crit}$. Consequently, there is no inherent advantage in operating in a small $\mathcal{Q}$, large $r$ regime for a beamsplitter QST process. This is in contrast to a three-mode mixing QST process, which, as we show below, can achieve a near-optimal Heisenberg-limited phase sensitivity in the small-$\mathcal{Q}$ regime. 

\section{Three-mode mixing QST process} \label{sec_single_mode_atom_int} 

\subsection{Analytical solution in small-QST regime} \label{sec_small_time_anal}
Although the dynamics of the three-mode mixing QST process cannot be analytically solved in general, an approximate solution does exist for times $t_1 \ll g^{-1}$ (equivalently, for $\mathcal{Q} \ll 1$). Defining $\tau \equiv  g t_1$, we can apply the improved Euler (i.e. Heun's) method \cite{Suli:2003} to Eqs.~(\ref{3modeEOM}) and obtain solutions valid to second order in $\tau$:
\begin{subequations}
\label{small_t_solns}
\begin{align}
	\hspace{-0.19 cm}\ahat_1(\tau) &=  \hat{a}_1 - i \tau \hat{b}^\dag \hat{a}_2 + \tfrac{\tau^2}{2}\hat{a}_1 (\hat{N}_{a_2} - \hat{N}_{b})  + \mathcal{O}(\tau^3), \\
	\hspace{-0.19 cm} \ahat_2(\tau) &=  \hat{a}_2 - i \tau \hat{b} \hat{a}_1 - \tfrac{\tau^2}{2}\hat{a}_2 (\hat{N}_{a_1} \hspace{-0.1 cm}+ \hat{N}_{b} +\hspace{-0.02 cm} 1) + \mathcal{O}(\tau^3), \\
	\hspace{-0.19 cm} \hat{b}(\tau)  &=  \hat{b} - i \tau \hat{a}_1^\dag \hat{a}_2 + \tfrac{\tau^2}{2}\hat{b} (\hat{N}_{a_2} - \hat{N}_{a_1}) + \mathcal{O}(\tau^3). 
\end{align}
\end{subequations}
For notational compactness, we have written $\hat{a}_1 \equiv \hat{a}_1(0)$, etc., $\hat{N}_{a_i} \equiv \hat{a}^\dag_i \hat{a}_i$, and $\hat{N}_{b} \equiv \hat{b}^\dag \hat{b}$. Note that an expansion to second order in $\tau$ is required in order to accommodate a non-zero QST efficiency, since $\mathcal{Q} = N_t \tau^2 + \mathcal{O}(\tau^3)$.

Noting that the average number of particles in mode $\hat{a}_1$ at time $\tau$ is
\begin{equation}
	\hspace{-0.031 cm}N_{a_1}(\tau) = N_t \big( 1 - \tau^2 [1 - \tfrac{\tau^2}{8}\left( 3 \cosh(2r) + 1 \right)] \sinh^2 r  \big),
\end{equation}
at the optimal operating point $\phi = \pi/2$, the derivative and variance of the information-recycled signal $\hat{\mathcal{S}}$ are
\begin{equation}
	\partial_\phi \langle \hat{\mathcal{S}} \rangle = \sqrt{N_t} \tau \left(N_{a_1}(\tau) - N_t \tau^2 \sinh^2 r \right)
\end{equation}
and
\begin{widetext}
\begin{align}
	\hspace{-0.05cm} V(\hat{\mathcal{S}}) 	&= e^{-2r}(1 - N_t \tau^2) N_{a_1}(\tau) \left[ 1 + N_t \tau^2 \left( e^{2r} + \tfrac{\tau^2}{4}(N_t - 1) - 1\right) \right] - N_t \tau^2 e^{-r} \sinh r \sqrt{N_t(1 - N_t \tau^2)N_{a_1}(\tau)} \notag \\
					& \times \left\{4 - \tfrac{\tau^2}{2}\left( 4N_t + 3 e^{-2r} - 1\right) + \tfrac{\tau^4}{16}\left[ 4 N_t \left( e^{2r} + 3 e^{-2r} - 2 \right) + 6 e^{3r} \cosh r + 9 e^{-2r} - 7 \right] \right\} \notag \\
					&+ N_t^2 \tau^2 \Big\{1 - 2 N_t \tau^2 e^{-r} \sinh r + \tfrac{\tau^4}{4}\left[ 1 + N_t^2 + N_t \left( 3 + 2 \left( 1 - 6 e^{-2r}\right)\sinh^2 r\right) + \tfrac{3}{2} \sinh^2(2r) \right] \notag \\
					& - \tfrac{\tau^6}{4}\sinh^2 r \left[ 2 + N_t^2 + \tfrac{N_t}{4}e^{-r}\left( 3 e^{-r}\left( 3 - 5 e^{-2r}\right) + e^r\left( 3 e^{2r} + 11\right)\right) + 6 \cosh(2r) \right] \notag \\
					& + \tfrac{\tau^8}{64}\sinh^2 r \left[ 2 N_t^2 \left( 3 \cosh(2r) + 1\right) + N_t\left( 15\left[ 2\cosh(2r) + \cosh(4r)\right] + 11 \right) + \tfrac{3}{2}\cosh^2 r \left( 35 \cosh(4r) + 13\right)\right] \Big\}, \label{VarS}
\end{align}
respectively. Substituting these quantities into the linear error propagation formula Eq.~(\ref{deltaphidef}), and performing a power series expansion around $\tau = 0$, gives the following expression for the phase sensitivity, valid at small times $\tau$:
\begin{align}
	\hspace{0 cm} \Delta \phi_{\mathcal{Q} \ll 1}(r, \tau) 	&= \frac{e^{-r}}{N_t \tau}\Bigg\{ 1 + \frac{3}{2} \tau^2 \sinh^2 r  + \frac{\tau^4}{8}\Big[ N_t\left( N_t + 3 - 2 e^{-2 r}\right) + 2\left( 5 \cosh(2r) - 6\right)\sinh^2r\Big]\Bigg\} + \mathcal{O}(\tau^5). \label{sens_tau}
\end{align}
\end{widetext}
In the $r \gtrsim 1$ regime, the first, second and third terms in Eq.~(\ref{sens_tau}) scale as $e^{-r}/\tau$, $\tau e^{r}$ and $(\tau e^r)^3$, respectively. 
$\Delta \phi$ will therefore approach infinity as $\tau \to 0$ unless we choose optimal $r = r_{\mathcal{Q} \ll 1}^\text{opt} \equiv \ln(\mathcal{C} / \tau)$ for some constant $\mathcal{C}$, in which case we obtain a minimum sensitivity of
\begin{equation}
	\Delta \phi_{\mathcal{Q} \ll 1}^{\min}(\mathcal{Q}) \approx \frac{\frac{5}{32} \mathcal{C}^3 + \frac{3}{8} \mathcal{C} + \frac{1 + \mathcal{Q}^2/8}{\mathcal{C}}}{N_t} + \mathcal{O}(\mathcal{Q}^{5/2}), \label{sens_to_min}
\end{equation}
where we have used $\tau = \sqrt{\mathcal{Q} / N_t}$, assumed the large $N_t$ limit, and ignored terms of order $\mathcal{O}(1 / N_t^2)$. Note that Eq.~(\ref{sens_to_min}) is a monotonically increasing function of $\mathcal{Q}$, and that this scaling is weak (strictly, $\mathcal{O}(\mathcal{Q}^2)$ to lowest order). Minimizing Eq.~(\ref{sens_to_min}) with respect to $\mathcal{C}$ gives $\mathcal{C} = \sqrt{\smash[b]{2(\sqrt{129} - 3)/15}}$, and therefore for sufficiently small $\mathcal{Q}$:
\begin{equation}
	\Delta \phi_\text{min}^{\mathcal{Q} \ll 1}  \approx \frac{49 + 3 \sqrt{129}}{(3 + \sqrt{129})^{3/2}}\frac{1}{N_t} \approx \frac{1.53}{N_t}. \label{min_sens}
\end{equation}
Surprisingly, our small-QST analytics predict that near-Heisenberg-limited sensitivities are achievable as $\mathcal{Q}$ approaches zero, with the caveat that $r$ must be very large. This is a key qualitative difference between the beamsplitter and three-mode mixing QST processes.

\subsection{Numerical Solution}

We complement our small-QST analytic solution - and quantitatively investigate the full $\mathcal{Q}$ parameter space - with numerical phase-space simulations of the three-mode mixing QST process. Common to all phase-space approaches is the conversion of the Heisenberg equations of motion to a partial differential equation (PDE) for a quasi-probability distribution \cite{Gardiner:2004b, Walls:2008}. Some approximation and/or mathematical ``trick'' is then employed to ensure this PDE takes the form of a Fokker-Planck equation (FPE) with a positive-definite diffusion matrix, as this can be efficiently simulated by a set of stochastic differential equations (SDEs). Below, we briefly introduce the truncated Wigner (TW) and positive-P (P$^+$) equations needed in order to simulate our interferometry scheme (see Sec.~\ref{sec_TW} and Sec.~\ref{sec_P}, respectively). The results of these simulations are presented in Sec.~\ref{sec_results}.

\subsubsection{Truncated Wigner (TW)} \label{sec_TW}

In this paper we have predominantly used the TW phase-space method \cite{Drummond:1993, Carter:1995, Steel:1998, Blakie:2008, Polkovnikov:2010}, which is based upon a PDE for the Wigner function. Typically, this PDE has third and higher-order derivatives that are truncated in order to give a FPE; although this is an uncontrolled approximation, it is frequently valid provided the occupation per mode is not too small for appreciable time periods \cite{Sinatra:2002, Johnsson:2013}. Here, the set of TW SDEs corresponding to Eqs.~(\ref{3modeEOM}) is
\begin{subequations}
\label{wigner_SDEs}
\begin{align}
	i \dot{\alpha}_1	&= g \alpha_2 \beta^*, \\
	i \dot{\alpha}_2	&= g \alpha_1 \beta, \\
	i \dot{\beta}   	&= g \alpha_2 \alpha_1^*,
\end{align} 
\end{subequations}
where we have made the correspondences $\ahat_{i}(t) \rightarrow \alpha_{i}(t)$ and  $\bhat(t) \rightarrow \beta(t)$. The initial conditions for these SDEs are randomly sampled from the Wigner distribution corresponding to the quantum state before the QST process (i.e. at time $t_0$) \cite{Olsen:2009}. Explicitly, $\hat{a}_1$ is in a coherent state of mean number $N_t$, $\hat{a}_2$ is in a vacuum state, and $\hat{b}$ is in a single-mode squeezed vacuum state. These correspond to the initial conditions
\begin{subequations}
\label{wigner_SDEs_initial}
\begin{align}
	\alpha_1(t_0) 		&= \sqrt{N_t} + \eta_{\alpha_1}, \\
	\alpha_{2}(t_0)		&= \eta_{\alpha_{2}}, \\
	\beta	(t_0)			&= \eta_{\beta} \cosh r  + \eta_{\beta}^* \sinh r,
\end{align}
\end{subequations}
where $\eta_i$ are complex, independent Gaussian noises satisfying $\overline{\eta_i} = 0$ and $\overline{\eta^*_i\eta_j} = \delta_{ij}/2$. 

The expectation value of any operator function $f$ is computed by averaging over solutions to Eqs.~(\ref{wigner_SDEs}) with initial conditions~(\ref{wigner_SDEs_initial}):
\begin{equation}
	\langle \{ f(\ahatd_1, \ahatd_2, \bhatd, \ahat_1, \ahat_2, \bhat ) \}_\mathrm{sym} \rangle = \overline{f\left(\alpha^*_1, \alpha^*_2, \beta^*, \alpha_1, \alpha_2, \beta \right)} \, ,
\end{equation}
where ``sym'' denotes symmetric ordering \cite{Walls:2008}, and the overline denotes the average of simulated trajectories. In order to compute the phase sensitivity with the information-recycled signal $\hat{\mathcal{S}}$ [Eq.~(\ref{eqn_info_recyc_sig})] at the optimal operating point, we made use of the following expressions:
\begin{subequations}
\begin{align}
	\langle \hat{N}_i(t_1) \rangle 		&= \overline{|\alpha_i(t_1)|^2} - 1/2, \\
	\langle \hat{J}_j(t_1) \rangle	&= \overline{\mathcal{J}_j}, \label{J_j}\\
	\langle \hat{J}_j(t_1)^2 \rangle	&= \overline{\mathcal{J}_j^2} - 1/8, \label{J_j_sq}\\
	\langle \hat{Y}_b(t_1) \rangle	&= -2 \overline{\text{Im}\left\{ \beta(t_1) \right\}}, \\
	\langle \hat{Y}_b(t_1)^2 \rangle	&= 4 \overline{\text{Im}\left\{ \beta(t_1) \right\}^2},
\end{align}
\end{subequations}
where $i = 1, 2$ and $j = x, z$, with $\mathcal{J}_x = [\alpha_1^*(t_1) \alpha_2(t_1) + \alpha_1(t_1) \alpha_2^*(t_1)]/2$ and $\mathcal{J}_z = [| \alpha_1(t_1)|^2 - | \alpha_1(t_2)|^2]/2$.

\subsubsection{Positive-P (P$^+$)} \label{sec_P}

In the regime where our low-QST analytics do not apply, we compared our TW numerical simulations to P$^+$ numerical simulations. The P$^+$ phase-space method expresses the evolution of the quantum state in terms of the Glauber-Sudarshan P representation \cite{Drummond:1980b, Gardiner:2004b, Walls:2008}. Unlike the Wigner representation, the evolution is guaranteed to be a FPE provided certain boundary terms are negligible (a condition that is valid for sufficiently short times \footnote{It is commonly claimed in the literature that P$^+$ simulations are exact. The use of `exact' here is perhaps misleading; provided there are no boundary terms the derivation of the FPE for the P$^+$ function is exact. However, in general boundary terms appear after some finite time, and the use of P$^+$ simulations in this regime is invalid. Furthermore, the SDEs associated with P$^+$ simulations in general contain multiplicative noise, which leads to a sampling error that grows rapidly in time. Fortunately, P$^+$ simulations typically diverge spectacularly in regimes where its application is invalid, although there are some notable exceptions \cite{Schack:1991, Gilchrist:1997, Dowling:2007}.}). However, a positive-definite diffusion matrix is only ensured by doubling the phase-space, which is effected by treating the complex amplitudes $\tilde{\beta}$ and $\tilde{\beta}^+ \equiv \tilde{\beta}^*$, for example, as independent variables. 

A set of P$^+$ SDEs corresponding to Eqs.~(\ref{3modeEOM}) is
\begin{subequations}
\label{eqs_P+}
\begin{align}
	\dot{\tilde{\alpha}}_1	&= -i g \tilde{\alpha}_2 \tilde{\beta}^+ + \sqrt{-i g} \tilde{\alpha}_2 w_1(t), \\
	\dot{\tilde{\alpha}}_2	&= -i g \tilde{\alpha}_1 \tilde{\beta}, \\
	\dot{\tilde{\beta}}   	&= -i g \tilde{\alpha}_2 \tilde{\alpha}_1^+ + \sqrt{-i g} w_1^*(t), \\
	\dot{\tilde{\alpha}}_1^+	&= i g \tilde{\alpha}_2^+ \tilde{\beta} + \sqrt{i g} \tilde{\alpha}_2^+ w_2(t), \\
	\dot{\tilde{\alpha}}_2^+	&= i g \tilde{\alpha}_1^+ \tilde{\beta}^+, \\
	\dot{\tilde{\beta}}^+  	&= i g \tilde{\alpha}_2^+ \tilde{\alpha}_1 + \sqrt{i g} w_2^*(t),
\end{align} 
\end{subequations}
where we have made the correspondences $\ahat_{i}(t) \rightarrow \tilde{\alpha}_{i}(t)$, $\ahat_{i}^\dag(t) \rightarrow \tilde{\alpha}_{i}^+(t)$, $\bhat(t) \rightarrow \tilde{\beta}(t)$, and $\bhat^\dag(t) \rightarrow \tilde{\beta}^+(t)$, and $w_i(t)$ are independent complex Wiener noises satisfying $\overline{w_i(t)} = 0$ and $\overline{w_i^*(t) w_j(t')} = \delta_{ij}\delta(t-t')$. Note that Eqs.~(\ref{eqs_P+}) can be interpreted as either Ito or Stratonovich SDEs, since the Stratonovich correction is zero. 

Our initial quantum state corresponds to the initial conditions \cite{Olsen:2009}
\begin{subequations}
\label{P+_IC}
\begin{align}
	\tilde{\alpha}_1(t_0) 	&= \tilde{\alpha}_1^+(t_0) = \sqrt{N_t}, \\
	\tilde{\alpha}_2(t_0) 	&= \tilde{\alpha}_2^+(t_0) = 0, \\
	\tilde{\beta}(t_0)	&= i \nu_-(r) n_1 - \nu_+(r) n_2 + \eta, \\
	\tilde{\beta}^+(t_0)	&= -i \nu_-(r) n_1 - \nu_+(r) n_2 - \eta^*,
\end{align}
\end{subequations} 
where $\nu_\pm(r) = \sqrt{ \exp(\pm r) \cosh(r) / 2}$, $\eta = (n_3 + i n_4) / \sqrt{2}$, and $n_i$ are real Gaussian random variables satisfying $\overline{n_i} = 0$ and $\overline{n_i n_j} = \delta_{ij}$.

Finally, normally-ordered operator expectations correspond to averages over the solutions to Eqs.~(\ref{eqs_P+}) with initial conditions Eqs.~(\ref{P+_IC}); for example
\begin{equation}
	\hspace{-0.065cm}\langle : \hspace{-0.1cm} f(\ahatd_1, \ahatd_2, \bhatd, \ahat_1, \ahat_2, \bhat ) \hspace{-0.1cm} : \rangle = \overline{f(\tilde{\alpha}_1^+, \tilde{\alpha}_2^+, \tilde{\beta}^+, \tilde{\alpha}_1, \tilde{\alpha}_2, \tilde{\beta})}, 
\end{equation}
where ``$: \, :$'' denotes normal ordering.

\subsubsection{Results} \label{sec_results}

\begin{figure}[!t]
\centering
\includegraphics[width=\columnwidth]{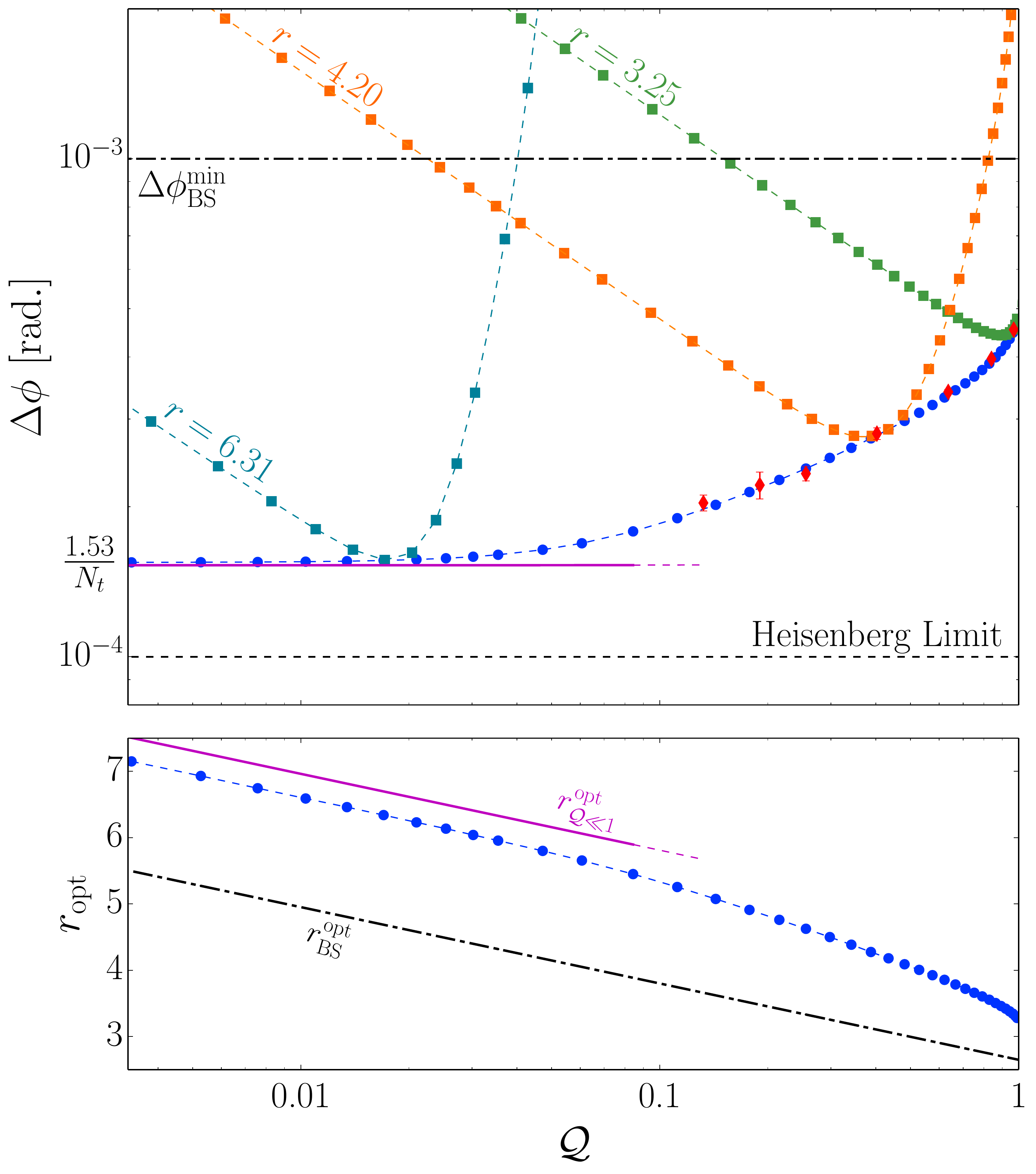}
\caption{ (Color online) Top panel: Phase sensitivity as a function of QST efficiency for a three-mode mixing QST process and an initial condensate of $N_t=10^4$ atoms. Points joined by dotted lines (simply to guide the eye) are TW simulations, whereas the disconnected red diamonds are P$^+$ simulations. More specifically, the three sets of square points (teal, orange, and green) show TW simulations of the phase sensitivity for fixed values of $r$, and the blue circles show the minimum possible phase sensitivity [i.e. for optimum $r = r_\text{opt}(\mathcal{Q})$] at particular values of $\mathcal{Q}$. The error bars on the P$^+$ simulations indicate twice the standard error, whereas the standard error in the TW simulations is less than the point width. The solid magenta curve is the small-QST analytic solution Eq.~(\ref{sens_to_min}), and the horizontal dashed line and dot-dashed line show the Heisenberg limit $1 / N_t$ and minimum possible phase sensitivity for a beamsplitter QST process $\Delta \phi_\text{BS}^\text{min} = N_t^{-3/4}$, respectively. Bottom panel: Optimal choice of $r$ for a fixed $\mathcal{Q}$ as predicted by a beamsplitter QST process [$r_\text{BS}^\text{opt}(\mathcal{Q}) = \ln(4N_t / \mathcal{Q}^2)/4$] (black dot-dashed curve), TW simulations (blue circles), and small-QST analytics for a three-mode mixing QST process [$r_{\mathcal{Q} \ll 1}^\text{opt}(\mathcal{Q}) = \ln(\mathcal{C}\sqrt{N_t / \mathcal{Q}})$] (solid magenta line). }  
\label{fig_TMM}
\end{figure}

The results of our TW and P$^+$ simulations are plotted in Fig.~\ref{fig_TMM}. There is excellent agreement between the TW and P$^+$ simulations in the large-QST regime, and between the TW simulations and the analytics of Sec.~\ref{sec_small_time_anal} in the low-QST regime, which give us confidence in the validity of our results. 

First, consider the three $\Delta \phi$ vs $\mathcal{Q}$ curves for fixed values of $r$. Clearly for $r < r_\text{crit}$ we see that the minimum phase sensitivity is always smaller than the minimum $\Delta \phi$ achievable with a beamsplitter QST process. This quantitative difference between the beamsplitter and three-mode mixing QST processes was noted in \cite{Szigeti:2014b}, although not explored in any detail. Here, our more comprehensive suite of results show also that for fixed $r > r_\text{crit}$, the minimum phase sensitivity occurs at a $\mathcal{Q} < 1$ - a feature also observed for the beamsplitter QST results in Fig.~\ref{fig_BS}. However, in contrast to the beamsplitter QST process, the minimum phase sensitivity \emph{decreases} as $r$ increases (and QST efficiency correspondingly decreases), and in fact is close to the absolute minimum $\approx 1.53 / N_t$ once the optimal $\mathcal{Q} \lesssim 5 \%$. This is shown very clearly by the blue circles in Fig.~\ref{fig_TMM}, which are TW simulations of the minimum phase sensitivity possible for a fixed $\mathcal{Q}$ (i.e. for the choice $r = r_\text{opt}(\mathcal{Q})$ - see bottom panel of Fig.~\ref{fig_TMM}). 

\subsection{Optimality with quantum Fisher information}

We now ask the question: is the information-recycled signal $\hat{\mathcal{S}}$ [Eq.~(\ref{eqn_info_recyc_sig})] the \emph{optimal} procedure for estimating the phase shift $\phi$? We answer this question using the quantum Fisher information $\mathcal{F}$, which places an absolute lower bound on the phase sensitivity $\Delta \phi \geq \Delta \phi_\text{QCRB} = 1/\sqrt{\mathcal{F}}$, called the quantum Cram{\'e}r-Rao bound (QCRB) \cite{Demkowicz-Dobrzanski:2014, Toth:2014}. Crucially, the QCRB applies irrespective of the choice of measurement and phase estimation procedure; it depends only on the input quantum state.  

In \cite{Lang:2013, Toth:2014}, it is shown that when a pure state at time $t_1$ forms the input to a lossless MZ interferometer, then the quantum Fisher information for estimating the relative phase shift between the two arms is simply $\mathcal{F} = 4 V(\hat{J}_y(t_1))$. As discussed in \cite{Haine:2014b}, since the initial state at time $t_0$ is pure, the three-mode mixing QST process is unitary, and we are permitting measurements on the photons transmitted after the QST process, then the state remains pure, and the quantum Fisher information for our hybrid atomic-photonic interferometer is the previously defined $\mathcal{F}$. Note that this would \emph{not} be true if we only allowed measurements on the two atomic modes; then we would have traced over the photon mode, the state would be mixed, and this simple expression for the quantum Fisher information no longer applies. 

Using Eqs.~(\ref{small_t_solns}), we obtained an analytic solution for the variance in $\hat{J}_y(t_1)$, and therefore the quantum Fisher information, valid in the small-$\tau$ (and therefore the small-QST) regime: 
\begin{widetext}
\begin{equation}
	\mathcal{F}_{\mathcal{Q}\ll 1}(r, \mathcal{Q}) = \mathcal{A}_1(r, \tau) N_t + \mathcal{A}_2(r, \tau) N_t^2 + \mathcal{A}_3(r, \tau) N_t^3 + \mathcal{O}(\tau^{10}), \label{Q_Fisher_info}
\end{equation}
where
\begin{subequations}
\begin{align}
	\mathcal{A}_1(r, \tau)	&= 1 + \frac{\tau^4}{16}\left[ 3 \cosh(4r)+1 \right] - \frac{\tau^6}{2}\left[ 3 \cosh(2r) +1\right]\sinh^2 r + \frac{3}{512}\tau^8 \left[ 35 \cosh(4r) +13\right] \sinh^2(2r), \\
	\mathcal{A}_2(r, \tau)		&= \tau^2 \left( e^{2r} - 1\right) - \frac{\tau^4}{4}\big[ 1 + 24 \cosh r \sinh^3 r - 7 \cosh(2r)+ 3 \cosh(4r) \big]  \notag \\
			& + \frac{\tau^6}{8}e^r \sinh^2 r \big[ \sinh r - 10 \cosh r + 9 \sinh(3r) + 6 \cosh(3r) \big] + \frac{\tau^8}{256}\big[ 8 - 23 \cosh(2r) + 15 \cosh(6r) \big], \\
	\mathcal{A}_3(r, \tau)	&= \frac{\tau^4}{4}\left( 1 - \tau^2 \sinh^2 r\right) + \frac{\tau^8}{32}\left[ 3 \cosh(2r) +1\right] \sinh^2 r.
\end{align}
\end{subequations}
\end{widetext}
We also numerically computed the quantum Fisher information from our TW simulations via Eq.~(\ref{J_j}) and Eq.~(\ref{J_j_sq}) for $j = y$ and $\mathcal{J}_y = i [\alpha_2^*(t_1) \alpha_1(t_1) - \alpha_2(t_1) \alpha_1^*(t_1)]/2$. 

\begin{figure}[t!]
\centering
\includegraphics[width=\columnwidth]{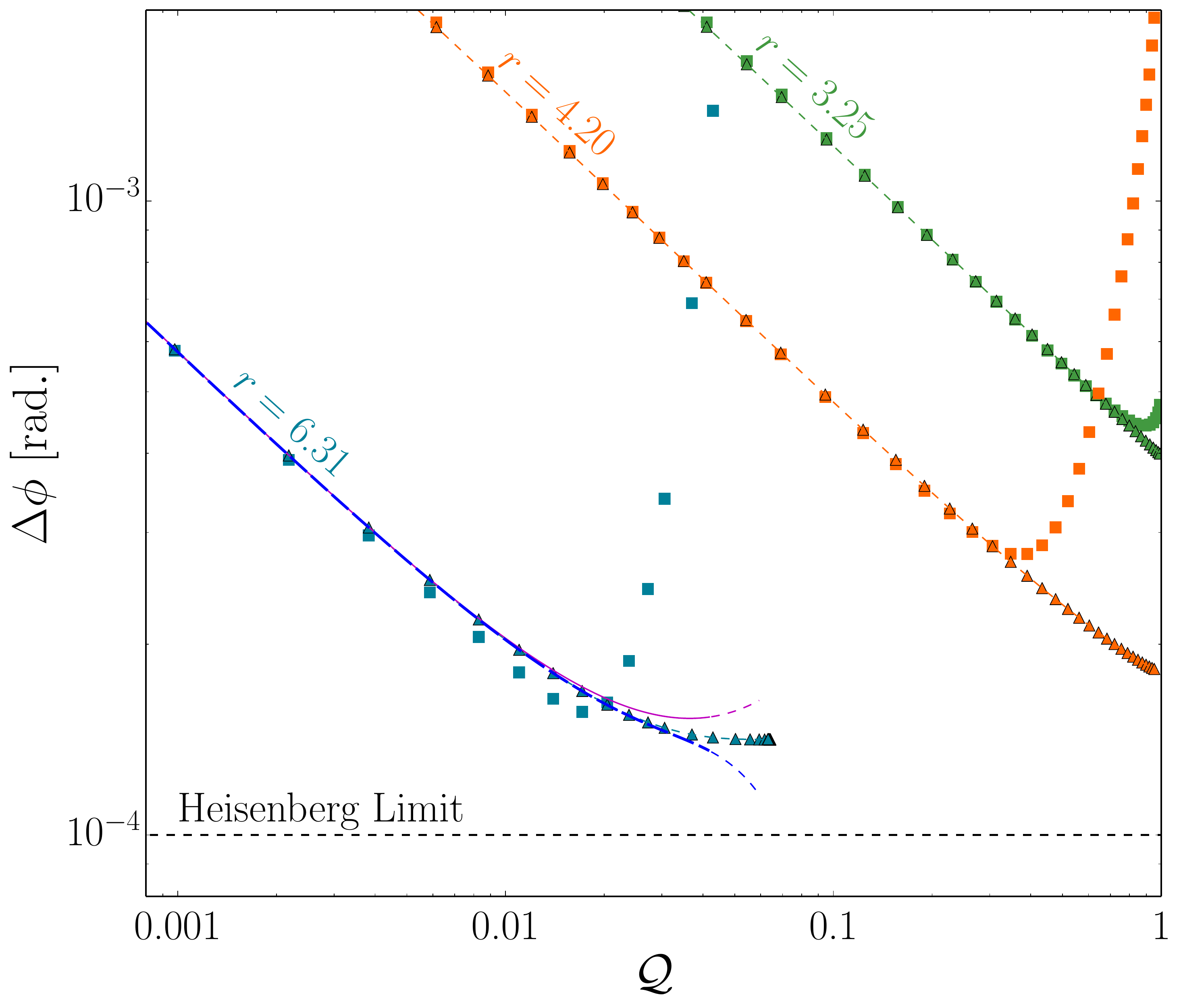}
\caption{ (Color online) Comparison of the QCRB and the phase sensitivity of the information-recycled signal for a three-mode mixing QST process and an initial condensate of $N_t = 10^4$ atoms. Points indicate the results of TW simulations; squares are $\Delta \phi$ calculated using $\hat{\mathcal{S}}$ [Eq.~(\ref{eqn_info_recyc_sig})], and triangles (joined by a dashed line to guide the eye) indicate the QCRB. The solid magenta and blue curves indicate the small-QST analytic curves Eq.~(\ref{sens_tau}) and $1 / \sqrt{\mathcal{F}_{\mathcal{Q} \ll 1}}$ [see Eq.~(\ref{Q_Fisher_info})], respectively, for $r = 6.31$. Here, there is a very good agreement between the TW simulations for the QCRB and the small-QST analytic solution, although there is a slight deviation between the TW simulation for the phase sensitivity of the information-recycled signal and Eq.~(\ref{sens_tau}). Since the standard error of all points is less than the point width, this discrepancy is caused by slight deviations from a physical Wigner function\footnote{A breakdown in the pure state expression for the quantum Fisher information, $\mathcal{F} = 4 V(\hat{J}_y(t_1))$, is not possible; since the QST dynamics are unitary, the Wigner function is guaranteed to remain pure, even under the TW approximation \cite{Corney:2015}.} due to the uncontrolled approximation associated with the TW simulation method.} 
\label{fig_fisher1}
\end{figure}

\begin{figure}[t!]
\centering
\includegraphics[width=\columnwidth]{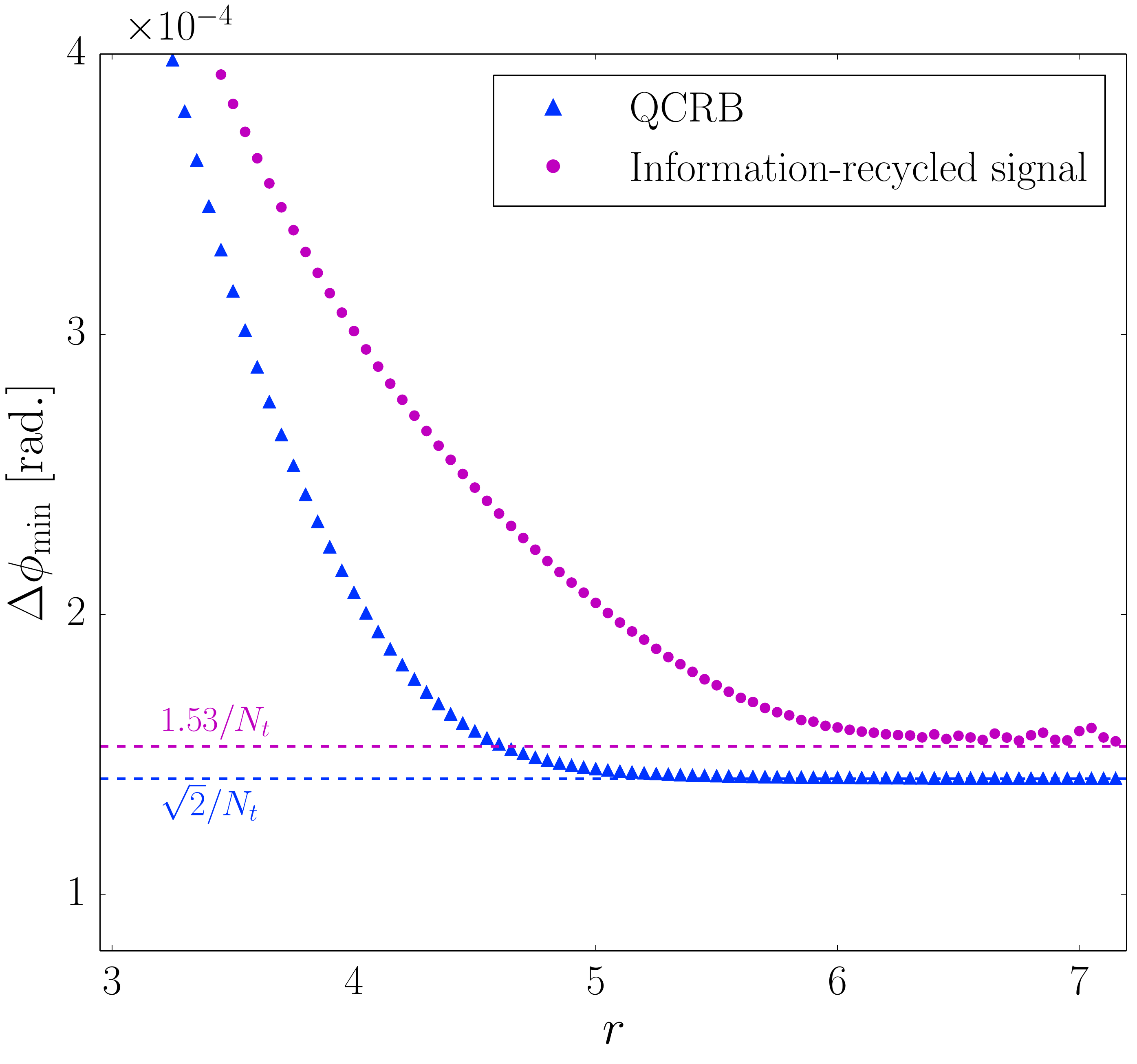}
\caption{ (Color online) Truncated Wigner simulations of the minimum phase sensitivity $\Delta \phi_\text{min}(r) \equiv \min_\mathcal{Q} \Delta \phi(r, \mathcal{Q})$ and minimum QCRB for a three-mode mixing QST process and initial condensate of $N_t = 10^4$ atoms. As $r$ gets large, the QCRB asymptotes to $\sqrt{2} / N_t$, which is only $\sim 7 \%$ lower than the minimum possible sensitivity $\Delta \phi_\text{min}^{\mathcal{Q} \ll 1} \approx 1.53 / N_t$ achievable with the information-recycled signal [see Eq.~(\ref{min_sens})]. The standard error in each simulation is less than the point width; consequently, the slight ``wiggles'' in the magenta points at large $r$ are systematic errors due to a breakdown in the uncontrolled approximation underlying the TW simulation method.}
\label{fig_fisher2}
\end{figure}

The phase sensitivity of the information-recycled signal is compared to the QCRB for three fixed values of $r$ in Fig.~\ref{fig_fisher1}. For $\mathcal{Q}$ less than some critical value (approximately equal to optimal $\mathcal{Q}$), the phase sensitivity saturates the QCRB. Close to the optimal $\mathcal{Q}$, where the phase sensitivity is minimized, the phase sensitivity increases, whereas the QCRB is non-increasing with increasing $\mathcal{Q}$. However, as shown in Fig.~\ref{fig_fisher2}, the minimum phase sensitivity is never more than a factor of $\sim 1.5$ larger than the QCRB. Furthermore, as $r$ increases, the minimum QCRB decreases to a minimum of $\sqrt{2} / N_t$, which is only $\sim 7\%$ lower than the minimum possible sensitivity achievable with the information-recycled signal Eq.~(\ref{eqn_info_recyc_sig}). We are therefore justified in calling our information-recycling scheme \emph{near-optimal}; currently an optimal phase-estimation scheme eludes us. 

Nevertheless, in the large $r$ regime the optimal phase estimation scheme requires a measurement of both the atoms at the interferometer output and the transmitted photons. Since $N_b = \sinh^2 r > N_t$, by definition it is impossible to achieve complete QST. If only the atoms are measured, then the quantum Fisher information depends on the reduced density matrix of the atoms alone. Here the quantum Fisher information is not simple to compute (it requires the diagonalization of the symmetric logarithmic derivative), although it is guaranteed to be less than or equal to $4 V(\hat{J}_y(t_1))$ \cite{Toth:2014}. Furthermore, since imperfect QST acts, to a first approximation, as a linear loss mechanism, Heisenberg scaling rapidly reverts to $1/ \sqrt{N_t}$ for even slight departures from $\mathcal{Q} = 1$ \cite{Demkowicz-Dobrzanski:2012}. This is consistent with the results of \cite{Szigeti:2014b}, which showed that without information recycling imperfect QST severely reduces the enhancement due to the squeezed light. Consequently, although information recycling is not strictly optimal, the optimal phase estimation procedure must incorporate information from both atomic and photonic measurements - as information recycling does - in regimes where only $\mathcal{Q} < 1$ is possible.  

\section{Conclusion} \label{Conclusion}
In most quantum technological applications that incorporate QST between two quantum systems, a larger QST efficiency is seen as more desirable. In this paper, we have provided a clear counter-example to this intuition. Specifically, we showed that if a hybrid atomic-photonic interferometer is enhanced by single-mode squeezed optical vacuum and information recycling (see Fig.~\ref{interferometer}), and the information-recycled signal Eq.~(\ref{eqn_info_recyc_sig}) is constructed, then for moderate to large levels of squeezing the best phase sensitivity requires an \emph{imperfect} QST efficiency. 

If the QST process is a beamsplitter, then this minimum phase sensitivity is $\Delta \phi_\text{BS}^\text{min} = N_t^{-3/4}$, for total atom number $N_t$. As shown in Fig.~\ref{fig_BS}, this minimum is obtained provided $r = r_\text{BS}^\text{opt} > r_\text{crit.}$, where $r_\text{BS}^\text{opt} \approx \ln(4 N_t / \mathcal{Q}^2)/4$ and $r_\text{crit} = \ln(4 N_t )/4$. In contrast, we showed that a three-mode mixing QST process yields a phase sensitivity close to the Heisenberg limit; counterintuitively, this is only achievable in the very small-QST, large squeezing regime (see Fig.~\ref{fig_TMM}). We compared this result to the QCRB, and showed that our scheme is nearly optimal (see Fig.~\ref{fig_fisher1} and Fig.~\ref{fig_fisher2}). In particular, the QCRB predicts a minimum phase sensitivity of $\sqrt{2} / N_t$, which is only $\sim 7\%$ lower than the minimum achievable with our information-recycled signal.

Unfortunately, the squeezing parameters required in order to realize these three-mode mixing QST dynamics are much larger than those achievable with current technology. We therefore see little prospect of an experimental observation of this effect in the near term. Nevertheless, the results of this paper are still of theoretical interest, as they demonstrate that information recycling is capable of exploiting metrologically useful correlations in regimes  where quantum metrology would n\"aively be deemed impossible. 

\section*{Acknowledgements}

We would like to acknowledge useful discussions with Joel Corney. Numerical simulations were performed using XMDS2 \cite{Dennis:2012} on the University of Queensland (UQ) School of Mathematics and Physics computer ``obelix'', with thanks to Ian Mortimer for computing support. SSS acknowledges the support of Ian P.~McCulloch and the Australian Research Council (ARC) Centre of Excellence for Engineered Quantum Systems (project no. CE110001013). SAH acknowledges the support of ARC Project DE130100575.

\bibliography{behnam_bib}

\end{document}